\documentclass[10pt,journal,letterpaper]{IEEEtran}
\usepackage{graphicx} 
\usepackage{hyperref}
\usepackage{amsmath}
\usepackage{amsfonts}
\usepackage{amssymb}
\usepackage{indentfirst}
\usepackage[dvipsnames]{xcolor}
\usepackage{pgfplots}
\usepackage{subcaption}
\usepackage{algorithm}
\usepackage[noend]{algpseudocode}
\usepackage{array}
\usepackage{cite}
\usepackage{doi}
\usetikzlibrary{calc}
\usetikzlibrary[shadows]
\usetikzlibrary{arrows.meta} 
\usetikzlibrary{decorations.pathreplacing,calligraphy}
\usetikzlibrary{positioning}
\usetikzlibrary{shapes.geometric}
\usepgfplotslibrary{fillbetween}

\definecolor{denim}{HTML}{3B638C}
\definecolor{darkred}{HTML}{840000}
\definecolor{virig}{HTML}{1fa187}
\definecolor{virig2}{HTML}{31688e}
\definecolor{lightblue}{HTML}{75bbfd}

\hypersetup{
	colorlinks=true,
	citecolor = blue,
	filecolor = blue,
	linkcolor = blue,
	urlcolor = blue,
	pdfstartview = FitH
}

\DeclareFontFamily{U}{wncy}{}
\DeclareFontShape{U}{wncy}{m}{n}{<->wncyr10}{}
\DeclareSymbolFont{mcy}{U}{wncy}{m}{n}
\DeclareMathSymbol{\Sh}{\mathord}{mcy}{"58}

\title{RSFQ All-Digital Programmable Multi-Tone Generator For Quantum Applications}
\author{Jo\~ao~Barbosa$^{\textbf{1,*}}$, Jack C. Brennan$^{\textbf{1}}$, Alessandro Casaburi$^{\textbf{2}}$, M.D. Hutchings$^{\textbf{3}}$, Alex Kirichenko$^{\textbf{3}}$, Oleg Mukhanov$^{\textbf{3}}$, Martin Weides$^{\textbf{1}}$ \\
\normalsize{\textbf{$^{\textbf{1}}$} James Watt School of Engineering, University of Glasgow, Scotland, United Kingdom} \\
\normalsize{{$^{\textbf{2}}$} Quantware, Elektronicaweg 10, 2628XG Delft, The Netherlands} \\
\normalsize{{$^{\textbf{3}}$} SEEQC, Inc., 150 Clearbrook Road, Elmsford, New York 10523, USA}
\thanks{$^{\textbf{*}}$j.barbosa.1@research.gla.ac.uk}
}

\begin{document}
\maketitle
\begin{abstract}
One of the most important and topical challenges of quantum circuits is their scalability. RSFQ technology is at the forefront of replacing current standard CMOS-based control architectures for a number of applications, including quantum computing and quantum sensor arrays. By condensing the control and readout to SFQ-based on-chip devices that are directly connected to the quantum systems, it is possible to minimise the total system overhead, improving scalability and integration.
In this work, we present a novel RSFQ device that generates multi tone digital signals, based on complex pulse train sequences using a Circular Shift Register (CSR) and a comb filter stage. We show that the frequency spectrum of the pulse trains is dependent on a preloaded pattern on the CSR, as well as on the delay line of the comb filter stage. By carefully selecting both the pattern and delay, the desired tones can be isolated and amplified as required. Finally, we propose some architectures where this device can be implemented to control and readout arrays of quantum devices, such as qubits and single photon detectors.
\end{abstract}

\section{Introduction}
Rapid Single Flux Quantum (RSFQ) \cite{OG1,RN9} electronics and their energy efficient version (ERSFQ) \cite{RN22,RN18} are well developed technologies that are now being considered and implemented for the control and readout of quantum circuits such as qubits \cite{RN3,RN30,RN31,RE1,RE2}, and quantum sensors \cite{RN15,RN36,RN37}.
As a result, many of the bulky standard room temperature electronics are being replaced by RSFQ circuits that can perform the same quantum circuit manipulation, but with potentially better scalability from decreased system overheads, better integration for cryogenic temperatures due to the much lower power dissipation, and faster operation (up to hundreds of GHz \cite{RN32}). Naturally, this leads to proposed architectures where a full manipulation of qubits and single photon detectors, such as Superconducting Nanowire Single Photon Detectors (SNSPDs), is done in a fully digital and, therefore, more scalable way \cite{RN2}.

One circuit of interest in this area is an arbitrary waveform generator (AWG), which at its core consists of a Digital-to-Analogue Converter (DAC), that can generate programmable arbitrary analogue signals. This is the quintessential device used in a myriad of control protocols, including standard microwave based single qubit gates \cite{RN39}  or time-domain measurements of RF-SNSPDs \cite{RN38}. In particular, AWGs are able, and widely used, to generate multi-tone signals for use in frequency division multiplexing (FDM) control architectures \cite{RN40,RN41}.
The RSFQ implementation of such a device usually consists of a digital pulse-based stage which encodes the analogue signal into a binary code that is then followed by a DAC stage, using voltage multipliers or SQUID stacks \cite{VM1,RN27,RN26,RN25}.
The only demonstration of multi-tone signal generation using SFQ-based circuits has been recently achieved for metrology applications using the same architecture type mentioned before, with a DAC at the output \cite{RN26}. Here we present an alternative way of generating programmable multi-tone signals with all-digital pulse based signals, which we name Digital Multi-Tone Generator (DMTG). 

Like the previously mentioned implementations of AWG-type RSFQ circuits, Circular Shift Registers (CSR) \cite{RN10,RN13,RN14,AUTO1} are at the core of this work too, since they can be used as pulse train generators and the type of pulse train is dependent on the initial pre-loaded data on the memory cells of the CSR. This property allows the creation of evenly and unevenly spaced patterns that can be controlled to generate the desired tones on the output signal. This is also an important property when trying to encode an analogue signal into a digital pulse train. The second important component is a Comb Filtering stage \cite{CF1} like those seen in Cascaded-Integrator-Comb filters, both in classical and RSFQ applications, used for decimation and interpolation filter devices \cite{RN28, RN23}. This component is also seen in some pulse multiplier circuits \cite{VM1}. In this work, comb filters are implemented to redistribute the power of the CSR output to frequencies that depend on the characteristics of the delay path. This improves the power of the desired tones and attenuates spurious tones if correctly tuned. A feedforward implementation is chosen to comply with RSFQ device restrictions, such as summation and multiplication of single SFQ pulses. 

\begin{figure*}[!ht]
\centering
\begin{tikzpicture} 
\node (sub1) at (-2.75,0) {\includegraphics[width=0.3\textwidth]{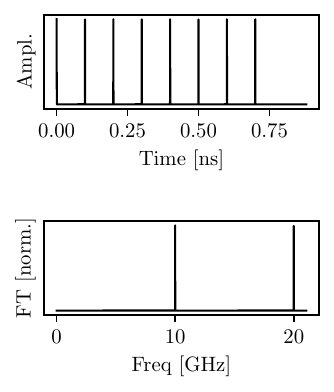}};
\node (sub2) at (3.,0) {\includegraphics[width=0.3\textwidth]{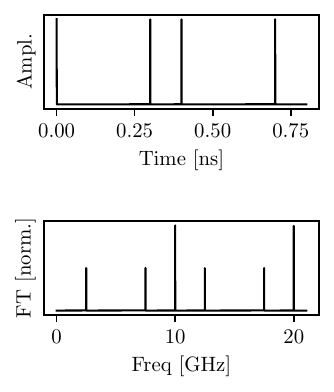}};
\node[text width=4cm,align=center] at (-2.5,3.4) {Pattern `1111 1111'};
\node[text width=4cm,align=center] at (3.25,3.4) {Pattern `1001 1001'};
\node (sub3) at (8.75,0) {\includegraphics[width=0.3\textwidth]{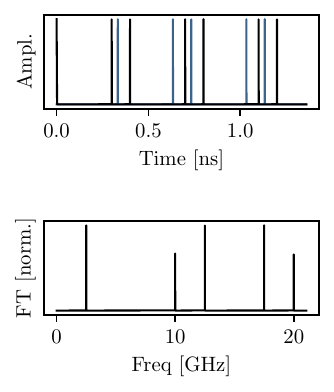}};
\node[text width=4cm,align=center] at (9.,3.52) {Pattern `1001 1001'\\ \& \textcolor{denim}{Comb Filter}};
\draw [stealth-stealth,color=denim](6.93,2.66)--(7.96,2.66);
\node[color=denim] at (7.26,2.48) {$\tau$};
\node[align=center] at ($(sub1.north west)+(0.3,0)$) {\textbf{(a)}};
\node[align=center] at ($(sub2.north west)+(0.3,0)$) {\textbf{(b)}};
\node[align=center] at ($(sub3.north west)+(0.3,0)$) {\textbf{(c)}};
\node[draw,rectangle,dashed,minimum width=2.2cm,minimum height=1.7cm,fill=gray,fill opacity=0.1] at ($(sub1.south west) + (2.05,2.14)$) (aux1) {};
\node[draw,rectangle,dashed,minimum width=2.2cm,minimum height=1.7cm,fill=gray,fill opacity=0.1] at ($(sub2.south west) + (2.05,2.14)$) (aux2) {};
\node[draw,rectangle,dashed,minimum width=2.2cm,minimum height=1.7cm,fill=gray,fill opacity=0.1] at ($(sub3.south west) + (2.05,2.14)$) (aux3) {};
\draw[thick,-{Triangle[open,length=6pt]}] ($(sub1.center)+(0.13,0.28)$) -- ($(sub1.center)+(0.13,-0.15)$);
\draw[thick,-{Triangle[open,length=6pt]}] ($(sub2.center)+(0.13,0.28)$) -- ($(sub2.center)+(0.13,-0.15)$);
\draw[thick,-{Triangle[open,length=6pt]}] ($(sub3.center)+(0.13,0.28)$) -- ($(sub3.center)+(0.13,-0.15)$);
\end{tikzpicture}
\caption{Example of pulse trains generated in an 8-bit CSR, with programmability focused on tones below $f_{\text{clk}}$ = 10GHz: (a) Evenly spaced pulse train generates an evenly spaced frequency spectrum (frequency division patterns). (b) Uneven spaced pulse train generates a richer spectrum with more tones. Pattern `1001 1001' generates tones at 2.5 and 7.5 GHz. All spectra from the CSR are periodic with period equal to $f_\text{clk}$. (c) Single stage comb filter applied on same pattern as (b), with delay $\tau$ = 0.33 ns, tunes the spectrum in a way to amplify the 2.5 GHz tone and cut off the 7.5 GHz tone. FT is normalised for the maximum power in each system, which corresponds to the pattern `1111 1111'. FT scale different between plots for visualisation purposes.}
\label{fig:fig1}
\end{figure*}
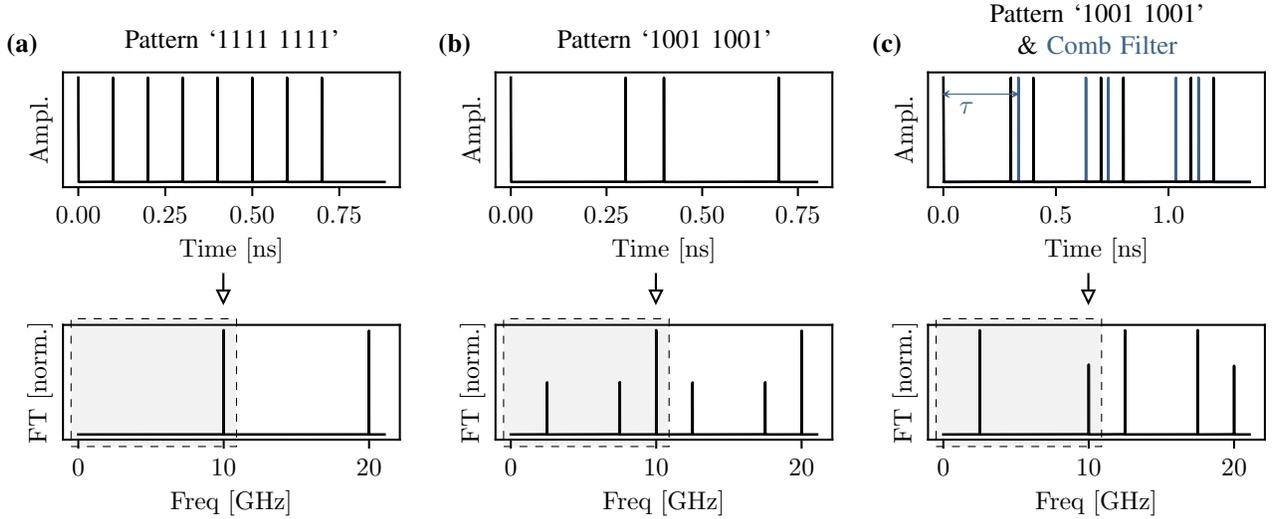
The remainder of this paper is organised as follows. Section \ref{ref:sec2} explains the mathematics of pulse trains and their frequency spectra, and how the use of CSR can modify them in a programmable way. Here an analysis of comb filters is also shown, as a way to improve the final spectrum of the signal. Section \ref{ref:sec3} then presents the proposed implementation using RSFQ basic components, with special emphasis on the CSR, the clocking network, and the comb filter stage. Section \ref{ref:sec4} shows the simulated results of a 8-bit DMTG using a superconductor circuit simulator, as well as a study of how SFQ pulse characteristics influence the final spectrum. In Section \ref{ref:sec5}, some applications envisioned for this device are presented. Finally conclusions are given in Section \ref{ref:sec6}.
\section{Multi-Tone Digital Signals with Pulse Trains} \label{ref:sec2}
Before introducing SFQ pulses in this work, this section will be concerned with the creation of multi-tone signals using ideal pulse trains. Later, we shall see that this approximation holds extremely well when dealing with SFQ pulses due to their sub picosecond time width.
\subsection{Spectral Analysis}
To start this analysis, we assume that a single SFQ pulse is modelled by an infinitely narrow pulse which takes the form of a Dirac delta function:
\begin{align}
    \delta(x)=
    \begin{cases}
    +\infty, &x=0 \\
    0, &x\neq 0 
    \end{cases}
\end{align}
A pulse train, like one generated by a simple SFQ-DC driver circuit clocked at a certain frequency $f_{\text{clk}}$, can also be written using Dirac functions, by infinitely summing them:
\begin{align}
\label{eq:eq2}
\Sh_\text{T}(x)  = \sum_{m=-\infty}^{+\infty} \delta(x-m\text{T})
\end{align}
where $\text{T}=1/f_{\text{clk}}$ and $\Sh$ represents the Dirac Comb function. These equally spaced pulse trains (with period T), have a Fourier Transform (FT) given by:
\begin{align}
    \mathcal{F}[{\Sh_{\text{T}}}(x)](f) = \int_{-\infty}^{+\infty} \Sh(x) e^{-2\pi j f x} \text{d}x = \frac{1}{\text{T}} \Sh_{1/\text{T}} (f)
\end{align}
This result shows that the frequency spectrum of an ideal and equally spaced pulse train, is itself an equally spaced pulse train with `period' given by 1/T, as shown in Fig. \hyperref[fig:fig1]{1(a)}. With these types of pulse trains, it is already possible to generate a multi-tone signal, composed of $f_{\text{clk}}$ and its harmonics, but the only way to control these tones is to change the frequency of the drive signal, which modifies all the tones in the signal. \par

Using a CSR, we can redefine the Dirac Comb function to include the pattern loaded in this circuit:
\begin{align}
    \tilde{\Sh}_{\text{T}}^{\text{N}}(x) = \sum_{m=-\infty}^{+\infty} \sum_{k=0}^{\text{N}-1} S_k \delta(x- k\text{T} - m \text{N} \text{T})
\end{align}
where the CSR is driven by a clock frequency $f_\text{clk}=1/T$, N is the number of bits of the CSR, and $S_k=\{0,1\}$ represents the bit (pattern) stored inside memory cell $k$ of the  CSR. The frequency spectrum of these unequally spaced pulse trains can be calculated in the same way and it is given by:
\begin{align} \label{eq:5}
    \mathcal{F}[\tilde{\Sh_{\text{T}}^{\text{N}}}(x)](f) = \frac{1}{\text{N}\text{T}} \Sh_{\frac{1}{\text{N}\text{T}}}(f) \cdot c(f)
\end{align}
where $c(f)$ is a modulation function given by the pattern in the CSR:
\begin{align}
    c(f)=\sum_{k=0}^{\text{N}-1}S_k e^{-2\pi j f k \text{T}}
\end{align}
This shows that the result of the CSR is to modulate a frequency pulse train with spacing equal to $f_\text{clk}/ \text{N}$ ($\Sh_{\frac{1}{\text{NT}}}$), depending on the pattern loaded on it ($S_k$). Using Euler's formula, the modulation function can be interpreted as a summation of sine waves, similar in concept to a Fourier Series, but limited in number and amplitudes. An example of this modulation is shown in Fig. \hyperref[fig:fig1]{1(b)}, with a pattern of `1001 1001', for a 8-bit CSR clocked at 10 GHz. With this, the tones at 2.5 and 7.5 GHz are generated at equal power, while keeping the clock frequency the same. The same tones are generated at \{2.5, 7.5\} + $n\cdot$10 GHz, since the FT keeps its periodicity of $f_\text{clk}$. \par

An important characteristic of the patterns and spectra generated by a CSR is the uniqueness of each one of them. We define a unique pattern for an N-bit CSR as a pattern that generates a frequency spectrum with a unique combination of tones, and a unique combination of relative powers between them, that no other pattern with the same bit length can generate. 

The total amount of possible patterns for an N-bit CSR is given by $2^\text{N} -1$. Since most of these are not unique, we define some rules to obtain a minimal set of patterns that can generate all others. To help us study the uniqueness of patterns, some definitions are useful: first, for a N-bit pattern, we define $S$ as the number of set bits, that is, the number of `ones' in the pattern. Secondly, we define a distance variable $\mathcal{D}$ as a set of distances (cyclical) between all set bits in the pattern, where we define a distance of 1 for two consecutive set bits. This set has a cardinality equal to $S$ and the sum of its values must equal the total number of bits N for a valid pattern.
For instance, the pattern `1001 1001' for an 8-bit CSR (see Fig. \hyperref[fig:fig1]{1(b)}), has a number of set bits $S=4$ and the distance set is $\mathcal{D}= \{ 3,1,3,1 \}$. 

The first rule that stands out to eliminate duplicate patterns is that any pattern with the same number of set bits $S$ and the same distance set $\mathcal{D}$ generates the same pattern. For instance, patterns `1000 0000' and `0100 0000' both have $S=1$ and $\mathcal{D}=\{8\}$. This type of bit shift results in a phase shift of the pulse train, but has no effect on the tones generated. This rule can be extended to include patterns whose distance set is a cyclic permutation (CP) of another (e.g. $\{1,3,1,3\}$ which is obtained by shifting $\{3,1,3,1\}$).

Secondly, infinitely long pulse trains exhibit time-reversal symmetry, meaning their spectra remain unchanged whether the pulse train is run from $t_0\rightarrow t_1$ or from $t_1 \rightarrow t_0$, where $t_1 > t_0$. This means that within a set of non-cyclical permutations (NCPs), we can remove further potential duplicate patterns by performing a `mirror' operation. For instance, taking a few periods of the pattern `1101 0000' with $\mathcal{D}=\{1,2,5\}$, and running it backwards ($-t$), it is possible to observe that the pattern `1100 0010' with $\tilde{\mathcal{D}}=\{1,5,2\}$ is obtained, both generating the same frequency spectrum, as confirmed through later simulations.

Heuristically, there is another rule that removes additional patterns from the unique set, despite not following completely our original definition of uniqueness. For each pattern obtained so far, there exists a pattern (defined as its dual) obtained by performing a bitwise NOT operation, that also generates the same tones, with the same power relative to each other, except the tones generated at \{n$\cdot f_\text{clk}$\} which have greater or lower power output depending on how many bits are set in the pattern. From the definition, we count every dual pattern as a unique pattern since the output power of these tones have a higher power than the original pattern, although the remaining tones powers are equal. For instance, the patterns `1000 0000' ($S=1$, $\mathcal{D}=\{8\}$) and `0111 1111' ($S=7$, $\overline{\mathcal{D}}=\{1,1,1,1,1,1,2 \}$) generate exactly the same tones, except the tones at \{n$\cdot f_\text{clk}$\} which have increased power for the latter (more sets bits correspond to higher total power at the output). By definition, every pair of dual patterns must have a sum of their set bits equal to N.

\begin{figure}[h]
    \centering
    \includegraphics[]{"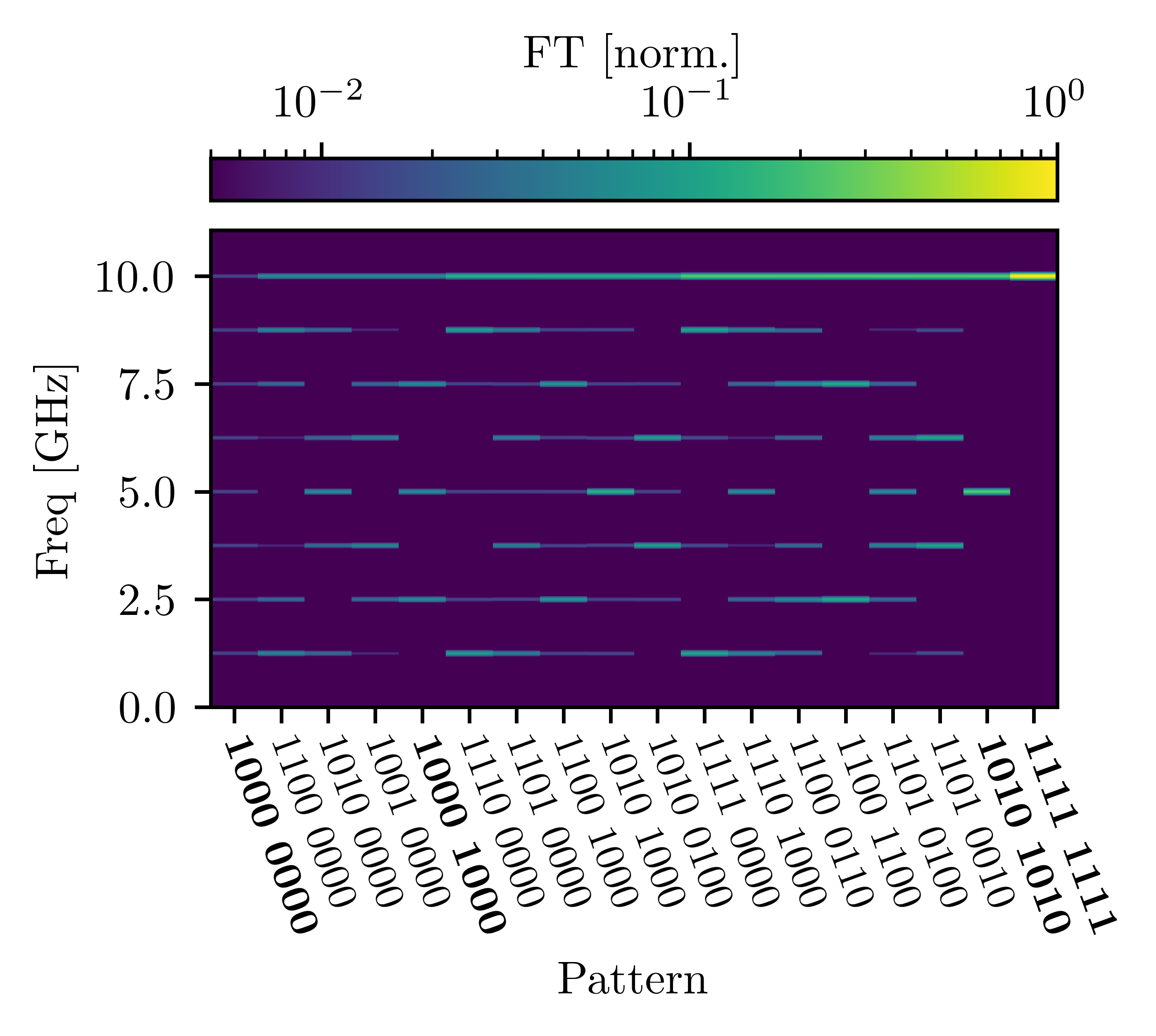"}
    \caption{Unique Patterns for an 8-bit CSR clocked at $f_{\text{clk}}= 10$ GHz. Uniqueness of each pattern defined by which tones it generates and the relative power between all of them. Bold patterns correspond to frequency division patterns (equally spaced in time and frequency). For each pattern shown (with exception of `1111 1111'), there exists a dual pattern that generates the same tone distribution but with the tone at $f_{\text{clk}}$ having increased power. Width of tones is artificially increased for visualisation purposes, since we are using Dirac delta functions. FT is normalised to the maximum output power of the system, corresponding to the pattern `1111 1111'.}
    \label{fig:fig2}
\end{figure}

With these dual patterns, one can obtain a smaller set of unique patterns below $f_\text{clk}$, since one needs only to look at unique patterns with set bits $S=\{1,2, \ldots , \text{N}/2\}$. These are shown for an 8-bit CSR in Fig. \ref{fig:fig2}, clocked at 10 GHz. The higher the number of set bits in the pattern, the higher the power spectral density. In a similar way, the more tones spread through the spectrum, the lower the spectral density for each one. This is easily seen with patterns `1000 0000' and `1111 1111', where the former has S=1 and all $f_\text{clk}$/N tones with their amplitude reduced by two orders of magnitude, when compared to the latter, with S=8 and only one tone.

These rules allow us to create a procedure to generate the unique patterns for any N-bit CSR, as shown in Algorithm \ref{algo1}. This procedure grows exponentially with the number of bits, since it requires various calculations of combinations and permutations. As expected, the number of unique patterns grows in a similar way, although slower than the original $2^\text{N}$. Two boundaries can be defined for a quick estimation of this number: the lower boundary is defined as the sum of all $s$-combinations of the set $\mathcal{S}=\{1,2,\ldots,\text{N}\}$ whose sum of elements equal N. The higher boundary is obtained from the latter, by adding the unique non-cyclic permutations (NCP) for each combination. The real number is then obtained by using procedure \ref{algo1} and dual patterns to eliminate some of the extra NCPs from the higher boundary that generate equal spectra.
\algrenewcommand{\algorithmiccomment}[1]{\hfill\textbf{//}\,#1}
\begin{algorithm}
    \caption{ Generating unique patterns on an N-bit CSR. }
    \begin{algorithmic}
        \State $s \gets 0$ \Comment{Number of set bits}
        \State $P_t \gets \{\}$ \Comment{Output set of unique patterns} 
        \While{$s \leq \text{N/2}$} 
            \State Empty set of dual patterns $\overline{P_t} \gets \{\}$
            \State Get all distance set combinations $D \gets {{\text{N}}\choose{s}}$
            \For {$D_j \text{ in } D$}
                \If{$\text{Sum}(D_j) = \text{N}$} 
                    \For{$d_i \text{ in } \text{NCPs}(D_j)$}
                    \If{$ \left( \text{CP}(d_i) \notin P_t \right) \land \left( \text{CP}(\tilde{d_i}) \notin P_t \right) \land \left( \text{CP}(\overline{d_i}) \notin \overline{P_t})\right) $} 
       
                        \State $P_t \gets d_i$
                        \State $\overline{P_t} \gets \overline{d_i}$

                    \EndIf
                    \EndFor
                \EndIf
            \EndFor
            \State $s \gets s + 1$
        \EndWhile
    \end{algorithmic}
    \label{algo1}
\end{algorithm}

Increasing the number of bits, N, in the shift register increases the frequency resolution since the base pulse train in the frequency space, defined in Eq. \eqref{eq:5}, is equally spaced by $1/\text{NT} = f_{\text{clk}}/\text{N}$. This means more tones can be generated within the same frequency range from 0 to $f_{\text{clk}}$. Another additional benefit of increasing the number of bits, is that for a long enough CSR, it is possible to simply overlap evenly spaced pulse trains of different frequencies and obtain a pattern with these frequencies only.


\subsection{Comb Filtering} 
Using a CSR with a pre loaded pattern increases the control of the tones generated with pulse trains. We can further improve it by using comb filtering stages as shown in this section. There are two types of comb filters: feedback and feedforward filters \cite{CF1}. These can be defined as two-port circuits, where a delayed version of the output or input signal (respectively) is added onto itself. Mathematically:
\begin{align} 
    &y(t) = x(t) + \alpha y(t-\tau), \text{   [Feedback]} \label{eq:7}\\
    &y(t) = x(t) + \alpha x(t-\tau), \text{   [Feedforward]} \label{eq:8}
\end{align}
where $x(t)$ is the input signal, $y(t)$ is the output signal, $\tau$ is the added delay and $\alpha$ is the scaling factor of the delayed signal. A block diagram of both types of filters is shown in Fig. \hyperref[fig:fig4]{4(b)}.
While the latter type has a potentially better amplitude response (specifically, it can target some tones more effectively due to the sharper frequency spectrum), its implementation using RSFQ circuits is not as straightforward, since the scaling factor needs be less than 1 to avoid instabilities in the filter output. To have the scaling factor less than 1, the top branch of the comb filter would need to have Josephson junctions with I$_\text{c}$R$_\text{N}$ product different from the rest of the circuit. Therefore, and as a first proof of concept, we look at the feedforward implementation and how it is used in the DMTG.\par 

Taking the Fourier Transform of Eq.\eqref{eq:8}, we obtain:
\begin{align}
    \mathcal{F}[y(t)](f) &= \mathcal{F}[x(t)](f) + \alpha \mathcal{F}[x(t-\tau)](f) \notag \\
    &= \left( 1+\alpha e^{-2\pi j f \tau} \right) \mathcal{F}[x(t)](f)
\end{align}
Like the previous section with the CSR pattern, this result shows that the modulation is achieved by summation of sine waves, but in this case, their frequency can be fine tuned by the delay of each filtering stage. There is, however, an exception to the usefulness of the comb filter, which happens when the delay matches the period spacing of the clock (1/$f_\text{clk}$). When such circumstances occur and because SFQ pulses cannot be added together (as in, on the same time instance, two pulses will not sum their amplitudes), the comb filter either produces no delayed pulse, so the spectrum will be exactly the same as it is seen at the end of the CSR, or it produces a delayed pulse which effectively changes the original pattern of the CSR into another pattern.

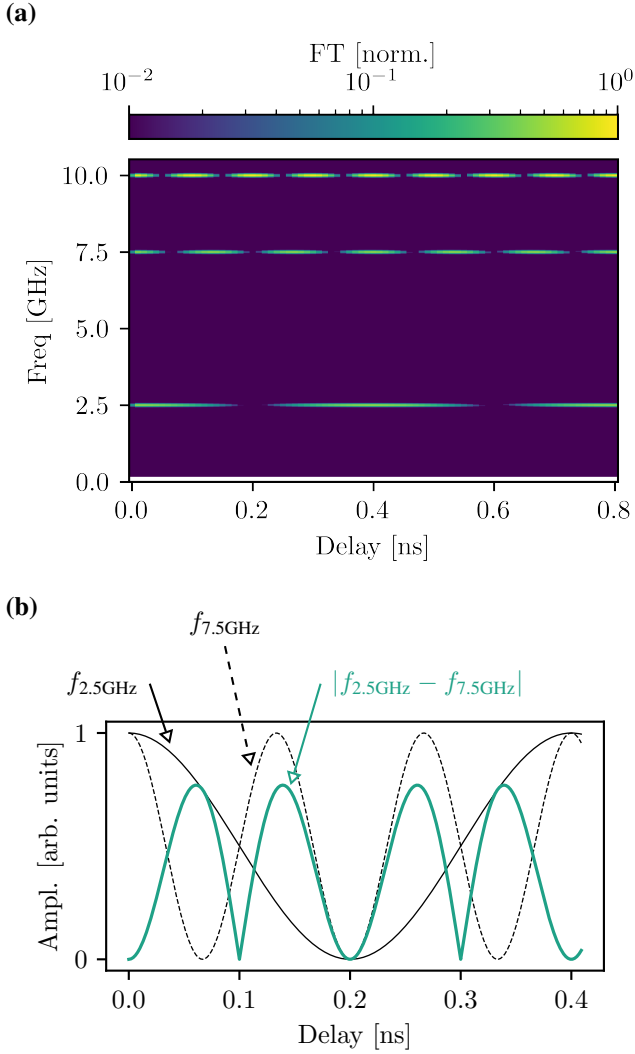
\begin{figure}[!h]
\begin{subfigure}{\columnwidth}
    \begin{tikzpicture}
    \node[anchor=center] (sub1) at (0,0) {\includegraphics[scale=1]{"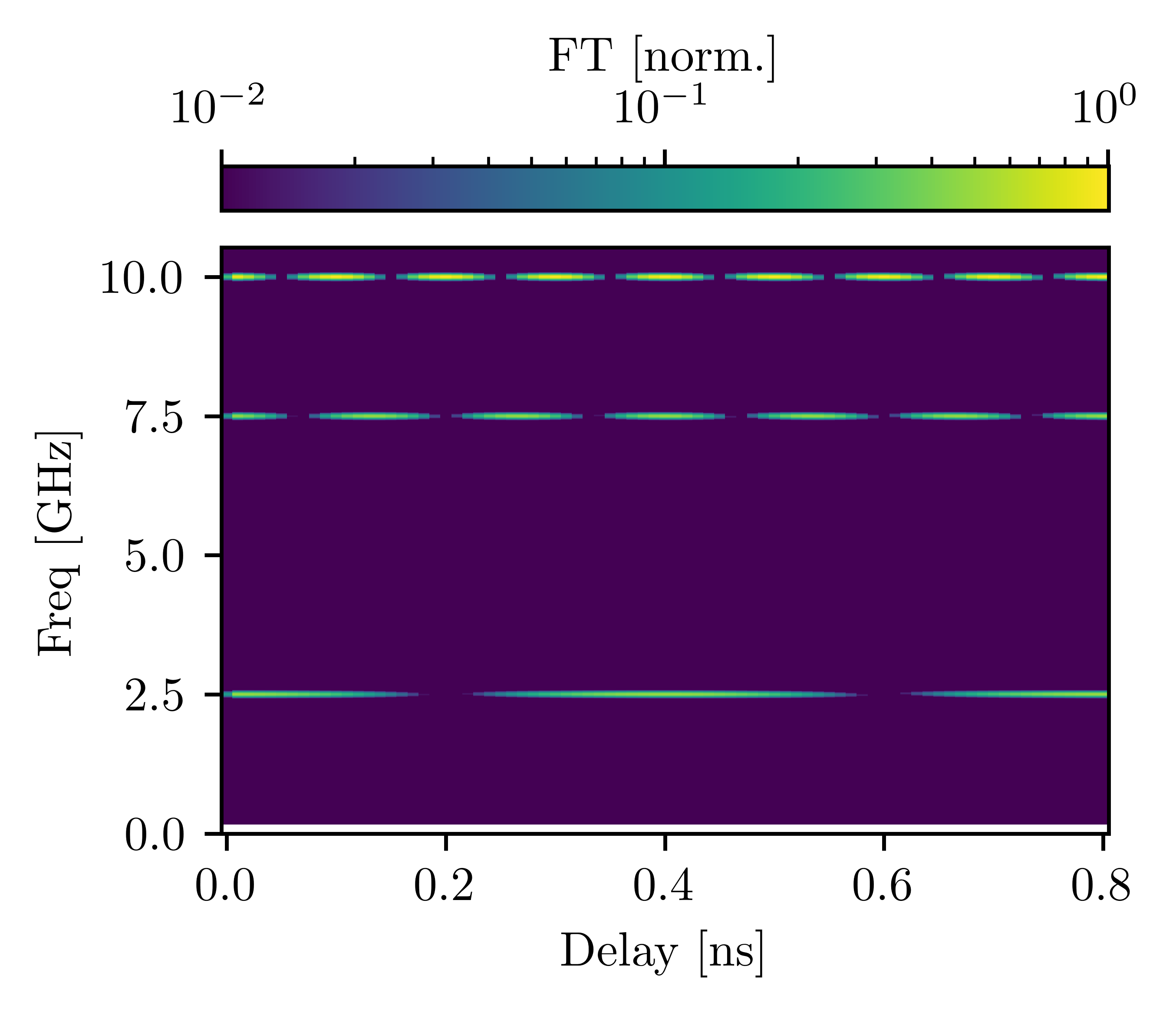"}};
    \node[align=center] at ($(sub1.north west)+(0.3,0)$) {\textbf{(a)}};
    \end{tikzpicture}
\end{subfigure}
\par
\begin{subfigure}{\columnwidth}
    \begin{tikzpicture}
        \node (sub1) at (0,0) {\includegraphics[]{"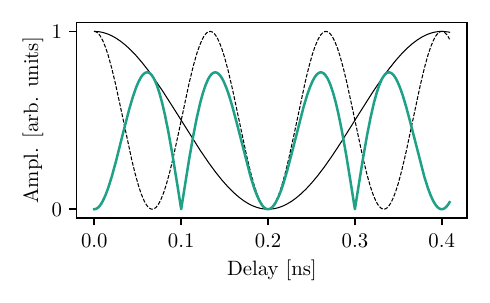"}};
        \draw[thick,color=virig,-{Triangle[open,length=5pt]}] (sub1.north) -- ++(-0.4,-1.35)      node[pos=0,right=0.02]{$ \left| f_{\text{2.5GHz}} - f_{\text{7.5GHz}} \right|$};
        \draw[thick,color=black,-{Triangle[open,length=5pt]}] ($(sub1.north west)+(2,0)$) -- ++(0.27,-0.8)      node[pos=0,left=0.02]{$f_{\text{2.5GHz}}$};
        \draw[thick,color=black,-{Triangle[open,length=5pt]},dashed] ($(sub1.north west)+(3,0.5)$) -- ++(0.35,-1.6)      node[pos=0,above=0.01]{$f_{\text{7.5GHz}}$};
        \node[align=center] at ($(sub1.north west)+(0.3,1)$) {\textbf{(b)}};
    \end{tikzpicture}
\end{subfigure}
\caption{Comb filter effect on frequency spectrum for the same pattern and CSR shown in Fig. \ref{fig:fig1} (b). (a) 2D view of spectrum versus comb filter delay. (b) Slice view of the two frequencies of interest (2.5 and 7.5 GHz) and a curve (green) illustrating the difference between these two latter, which is utilised to adjust the two tones in relation to each other. In this example, the largest difference is achieved for a delay $\sim$ 0.066 ns, or 0.333 ns, where the amplitude of the 7.5 GHz tone is reduced to zero; or 0.133 and 0.266 ns, where the latter tone is amplified and the 2.5 GHz tone reduced, but not removed.}
\label{fig:fig3}
\end{figure}

Fig. \ref{fig:fig3} shows an example of the effect of using a feedforward comb filter stage on a 8-bit CSR with a preloaded pattern of `1001 1001'. As the delay is tuned from 0 to N$/ f_\text{clk}$, the tones at 2.5, 7.5 and 10 GHz are periodically tuned from 0 to a maximum amplitude. A cross section of the 2D plot can also be seen, where the periodic behaviour of the tones at 2.5 and 7.5 GHz is plotted, together with a curve showing the distance between these two tones. By selecting the correct delay, we can tune both tones to be at a maximum of separation (for example to suppress the 7.5 GHz tone, with delay 0.33ns) or a minimum of separation, which for this pattern corresponds to both tones at equal strength. Fig. \hyperref[fig:fig1]{1(c)} shows the result of selecting a delay equal to 0.33 ns on the output of the CSR. As predicted, the tone at 7.5 GHz is suppressed.


\section{RSFQ Implementation} \label{ref:sec3}

\subsection{Circular Shift Registers}
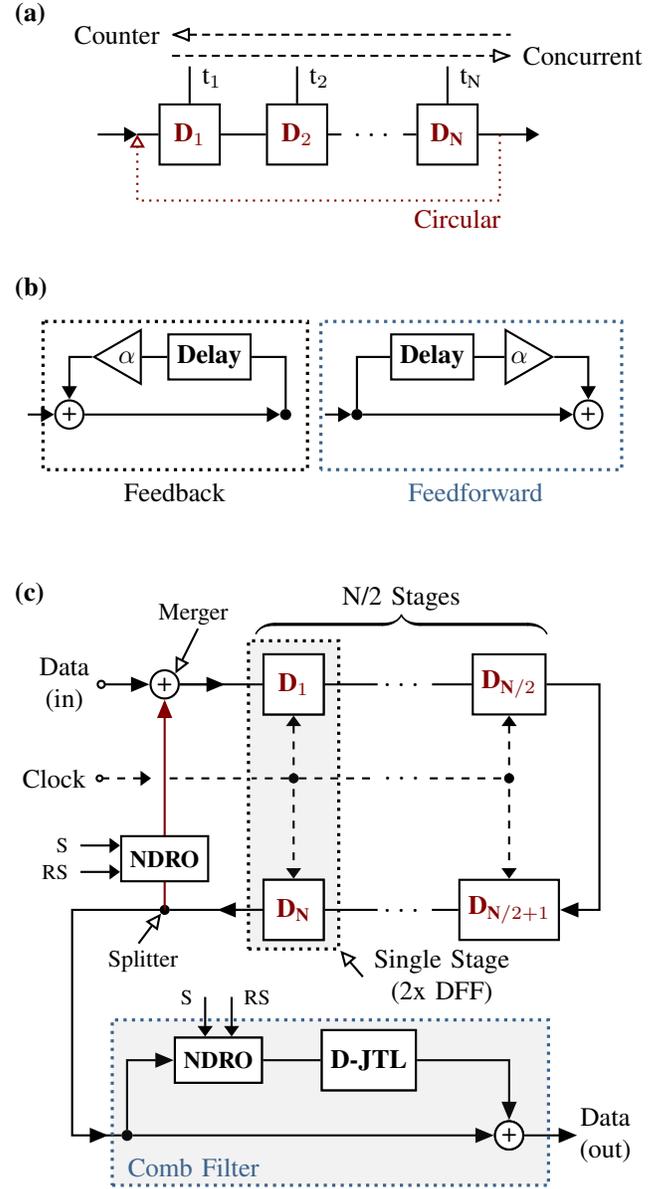
\begin{figure}[!ht]
\begin{subfigure}{0.48\textwidth}
    \begin{tikzpicture}
        \node[thick] at (\textwidth,0) (e1) {};
\node[thick] at (0,0) (e0) {};

\node[draw, thick, rectangle, minimum width=0.8cm, minimum height=0.8cm, fill=white] at (\textwidth*0.31,0) (d1) {\color{darkred}\textbf{D$_1$}};
\node[draw, thick, rectangle, minimum width=0.8cm, minimum height=0.8cm, fill=white,right=0.6 of d1] (d2) {\color{darkred}\textbf{D$_2$}};
\node[thick,,right=0.2 of d2]  (cd1) {\large $\ldots$};
\node[draw, thick, rectangle, minimum width=0.8cm, minimum height=0.8cm, fill=white,,right=2.6 of d1] (d3) {\color{darkred}\textbf{D$_\text{N}$}};

\draw[thick] (d1.north) -- +(0,14pt) node[below right=-3pt and 1pt,align=center] (n1) {t$_1$};
\draw[thick] (d2.north) -- +(0,14pt) node[below right=-3pt and 1pt,align=center] (n2) {t$_2$};
\draw[thick] (d3.north) -- +(0,14pt) node[below right=-3pt and 1pt,align=center] (n3) {t$_\text{N}$};

\draw[thick] (d1.east) -- (d2.west);
\draw[thick] (d2.east) -- ++(5pt,0);
\draw[thick] (d3.west) -- ++(-5pt,0);

\draw[thick] (d1.west) -| ++(-8pt,0) node[] (a) {};
\draw[thick] (d3.east) -| ++(8pt,0) node[] (b) {};
\draw[thick,-{Triangle[open,length=5pt]},dotted,color=darkred] (b.center) -- ++(0,-25pt) -| (a.center);

\draw[thick,densely dashed,-{Latex[length=8pt,round,open]}] ($(n1.north) + (-15pt,1pt)$) -- ($(n3.north) + (15pt,1pt)$) node[pos=1,right=1pt]{Concurrent};
\draw[thick,densely dashed,{Latex[length=8pt,round,open]}-] ($(n1.north) + (-15pt,9pt)$) -- ($(n3.north) + (15pt,9pt)$) node[pos=0,left=1pt]{Counter};

\draw[thick,-{Triangle[length=5pt]}] (b.center) -- ++(15pt,0);
\draw[thick,{Triangle[length=5pt]}-] (a.center) -- ++(-15pt,0);

\node[align=center] at ($(b.center)+(-16pt,-32pt)$) (csr) {\color{darkred}{Circular}};

\node[align=left,above right=1.2cm and 0.12cm of e0] {\textbf{(a)}};
    \end{tikzpicture}
\end{subfigure}
\par\bigskip
\begin{subfigure}{0.49\textwidth}
    \begin{tikzpicture}
\node[thick] at (\textwidth,0) (e1) {};
\node[thick] at (0,0) (e0) {};

\node[draw, thick, rectangle, minimum width=0.8cm, minimum height=0.6cm, fill=white] at (\textwidth/3,0) (d1) {\textbf{Delay}};
\node[draw, thick, rectangle, minimum width=0.8cm, minimum height=0.6cm, fill=white] at (2\textwidth/3,0) (d2) {\textbf{Delay}};

\node[draw, thick, isosceles triangle,isosceles triangle apex angle=60,rotate=180, minimum size=1pt,inner sep=2pt, fill=white,left=1 of d1] (t1) {\rotatebox{180}{\textbf{$\alpha$}}};

\node[draw, thick, isosceles triangle,isosceles triangle apex angle=60, minimum size=1pt,inner sep=2pt, fill=white,right=0.4 of d2] (t2) {\textbf{$\alpha$}};

\draw[thick] (d1.west) -- (t1.west);
\draw[thick] (d2.east) -- (t2.west);

\node[draw,thick,fill, circle,minimum size=1pt, inner sep=1.2pt, below right =0.4 and 0.4 of d1](s1) {};
\node[draw, thick, circle, minimum size=1pt, inner sep=1.1pt,left=2.6 of s1] (m1) {+};

\node[draw,thick,fill, circle,minimum size=1pt, inner sep=1.2pt, right=0.8 of s1](s2) {};
\node[draw, thick, circle, minimum size=1pt, inner sep=1.1pt, right=2.8 of s2] (m2) {+};

\draw[thick,-{Triangle[length=5pt]}] (t1.east) -| (m1.north);
\draw[thick] (d1.east) -| (s1.north);
\draw[thick,-{Triangle[length=5pt]}] (m1.east) -- (s1.west);

\draw[thick,-{Triangle[length=5pt]}] (t2.east) -| (m2.north);
\draw[thick] (d2.west) -| (s2.north);
\draw[thick,-{Triangle[length=5pt]}] (s2.east) -- (m2.west);

\draw[thick,{Triangle[length=5pt]}-] (m1.west) -- ++(-10pt,0);
\draw[thick,{Triangle[length=5pt]}-] (s2.west) -- ++(-10pt,0);

\node[align=left,above right=0.5cm and 0.12cm of e0] (l1) {\textbf{(b)}};
\node[below right=1pt and 1pt of l1.south] (ll1) {};
\draw[very thick, dotted, rectangle] ($(ll1.center)+(0,-0pt)$) rectangle (\textwidth*0.48,-1.5) {};
\draw[very thick, dotted, rectangle,color=denim] (\textwidth/2,-1.5) rectangle (\textwidth*0.95, 52 |- ll1.center) {};

\node[thick,align=center] at (2.5,-1.8) (r1) {Feedback};
\node[thick,align=center,color=denim] at (6.5,-1.8) (r1) {Feedforward};
    \end{tikzpicture}
\end{subfigure}
\par\bigskip\bigskip
\begin{subfigure}{0.5\textwidth}
    \begin{tikzpicture}
\node[thick] at (\textwidth,0) (e1) {};
\node[thick] at (0,0) (e0) {};

\node[draw, thick, rectangle, minimum width=0.8cm, minimum height=0.8cm, fill=white] at (\textwidth*0.45,0) (d1) {\color{darkred}\textbf{D$_1$}};
\node[draw, thick, rectangle, minimum width=0.8cm, minimum height=0.8cm, fill=white] at (\textwidth*0.45,-3) (d2) {\color{darkred}\textbf{D$_{\text{N}}$}};

\draw[thick,dashed,{Triangle[length=4pt]}-{Triangle[length=4pt]}] (d1.south) -- (d2.north);
\node[draw, circle, fill,minimum size=1pt, inner sep=1.2pt] at (\textwidth*0.45,-1.25) (n1) {};
\draw[thick] (d1.east) -- +(20pt,0);

\draw[thick,dashed] (n1.west) -- +(-14pt,0);
\draw[thick,dashed] ($(n1.west) + (-18pt,0)$) -- +(-28pt,0) node[pos=0.7,above=1pt]{};
\draw[thick,dashed,{Triangle[length=4pt]}-{Circle[open,length=3]}] ($(n1.west) + (-52pt,0)$) -- +(-21pt,0) node[pos=1,align=center,anchor=east]{Clock};
\draw[thick,dashed] (n1.east) -- +(30pt,0);

\draw[thick] (d2.east) -- +(20pt,0);

\node[draw,thick, circle,minimum size=1pt, inner sep=1.1pt] at ($(d1.west)+(-1.3,0)$) (n3) {+};
\draw[thick,{Triangle[open,length=6pt]}-] (n3.north east) -- ++(0.3,0.6) node[pos=0.8,above=1pt]{\small{Merger}};
\node[draw, circle, fill,minimum size=1pt, inner sep=1.2pt] at ($(d2.west)+(-1.3,0)$) (n4) {};
\draw[thick,{Triangle[open,length=6pt]}-] (n4.south west) -- ++(-0.3,-0.4) node[pos=0.8,below=1pt]{\small{Splitter}};
\draw[thick,-{Triangle[length=7pt]}] (n3.east) -- +(+0.6,0);
\draw[thick] (n3.east) -- (d1.west) ;
\draw[thick] (n4.east) -- (d2.west);
\draw[thick,-{Triangle[length=7pt]}] (d2.west) -- +(-0.6,0);

\draw[thick,-{Triangle[length=7pt]},color=darkred] (n4.center) -- (n3.south) ;

\draw[thick,{Triangle[length=7pt]}-{Circle[open,length=3.5]}] (n3.west) -- +(-20pt,0) node[pos=0.9,left=1pt,align=center] {Data \\ (in)};

\node[thick] at ($(n1.east)+(39pt,0)$) (cn1) {\large \textbf{$\ldots$}};
\node[thick] at ($(d1.east)+(29pt,0)$) (cd1) {\large $\ldots$};
\node[thick] at ($(d2.east)+(29pt,0)$) (cd2) {\large $\ldots$};

\draw[thick,dashed] (cn1.east) -- +(30pt,0);
\node[draw, circle, fill,minimum size=1pt, inner sep=1.2pt] at ($(cn1.east)+(30pt,0)$) (n2) {};

\node[draw, thick, rectangle, minimum width=0.8cm, minimum height=0.8cm, fill=white] at ($(n2.center)+(0,1.25)$) (d3) {\color{darkred}\textbf{D$_{\text{N}/2}$}};
\draw[thick] (d3.west) -- +(-15pt,0);

\node[draw, thick, rectangle, minimum width=0.8cm, minimum height=0.8cm, fill=white] at (d2 -| d3) (d4) {\color{darkred}\textbf{D$_{\text{N}/2+1}$}};
\draw[thick] (d4.west) -- +(-10pt,0);
\draw[thick,dashed,{Triangle[length=4pt]}-{Triangle[length=4pt]}] (d3.south) -- (d4.north);
\draw[thick, -{Triangle[length=7pt]}] (d3.east) -- +(20pt,0)|- (d4.east);

\node[draw, dotted, very thick, rectangle, minimum width=1.2cm, minimum height=4.1cm, fill=gray, fill opacity=0.1] at ($(n1.center) + (0,-0.22)$) (l1) {};
\draw[thick, {Triangle[open,length=6pt]}-] (l1.south east) -- +(0.3,-0.35) node[pos=1,align=center,right=0.5pt]{Single Stage\\(2x DFF)};

\draw[very thick,decorate, decoration={calligraphic brace, raise=5pt,amplitude=7pt}] ($(d1.center)+(-0.5,0.5)$) -- ($(d3.center)+(0.5,0.5)$) node[pos=0.5,above=10pt]{N/2 Stages};

\node[draw, thick, rectangle, minimum width=0.8cm, minimum height=0.8cm, fill=white] at ($(d2.center)+(1,-2)$) (d5) {\textbf{D-JTL}};
\node[draw, thick, rectangle, minimum width=0.8cm, minimum height=0.6cm, fill=white] at ($(d2.center)+(-1,-2)$) (ndro2) {\small{\textbf{NDRO}}};
\node[draw,thick,fill, circle,minimum size=1pt, inner sep=1.2pt] at ($(n4.center)+(-0.5,-3)$) (n5) {};
\node[draw, thick, circle, minimum size=1pt, inner sep=1.1pt] at (d4.center |- n5.center) (n6) {+};

\draw[thick,-{Triangle[length=7pt]}] (n4.center) -- ++(-35pt,0) |- ($(n5.west) + (-5pt,0)$);
\draw[thick] ($(n5.west) + (-12pt,0)$) -- (n5.west);
\draw[thick,-{Triangle[length=7pt]}] (n5.north) |- (ndro2.west);
\draw[thick] (ndro2.east) -- (d5.west);
\draw[thick,-{Triangle[length=7pt]}] (n5.east) -- (n6.west);
\draw[thick,-{Triangle[length=7pt]}] (d5.east) -| (n6.north);
\draw[thick,-{Triangle[length=7pt]}] (n6.east) -- ++(20pt,0) node[pos=0.85,align=center,right=0.2pt]{Data\\(out)};

\node[draw, dotted, very thick, rectangle, minimum width=5.8cm, minimum height=2.2cm,color=denim,fill=gray, fill opacity =0.1] at ($(d5.south east)+(-1.15,-0.15)$) (l2) {};
\node[draw, thick, rectangle, minimum width=0.8cm, minimum height=0.8cm, fill=white] at ($(d2.center)+(1,-2)$) (d51) {\textbf{D-JTL}};
\node[draw, thick, rectangle, minimum width=0.8cm, minimum height=0.6cm, fill=white] at ($(d2.center)+(-1,-2)$) (ndro222) {\small{\textbf{NDRO}}};
\node[draw, thick, circle, minimum size=1pt, inner sep=1.1pt,fill=white] at (d4.center |- n5.center) (n6) {+};
\node[draw, thick, rectangle, minimum width=0.8cm, minimum height=0.8cm, fill=white] at (\textwidth*0.45,0) (d111) {\color{darkred}\textbf{D$_1$}};
\node[draw, thick, rectangle, minimum width=0.8cm, minimum height=0.8cm, fill=white] at (\textwidth*0.45,-3) (d222) {\color{darkred}\textbf{D$_{\text{N}}$}};

\node[thick,color=denim,align=left,anchor=west] at ($(l2.south west) +(0.1,0.255)$) (cb) {Comb Filter};

\node[draw, thick, rectangle, minimum width=0.8cm, minimum height=0.6cm, fill=white, above=0.3 of n4] (ndro) {\textbf{\small{NDRO}}};

\draw[thick,{Triangle[length=4pt]}-] ($(ndro.west)+(0,5pt)$) -- ++(-15pt,0) node[pos=1,left=1pt,align=center]{\footnotesize{S}};
\draw[thick,{Triangle[length=4pt]}-] ($(ndro.west)+(0,-5pt)$) -- ++(-15pt,0) node[pos=1,left=1pt,align=left]{\footnotesize{RS}};
\draw[thick,{Triangle[length=4pt]}-] ($(ndro2.north)+(-5pt,0pt)$) -- ++(0,15pt) node[pos=1,left=1pt,align=center]{\footnotesize{S}};
\draw[thick,{Triangle[length=4pt]}-] ($(ndro2.north)+(5pt,0pt)$) -- ++(0,15pt) node[pos=1,right=1pt,align=center]{\footnotesize{RS}};

\node[align=left,above right=0.80cm and 0.12cm of e0] {\textbf{(c)}};

\end{tikzpicture}
\end{subfigure}
\caption{(a) Diagram of N-bit linear shift register being clocked in a concurrent or counter flow. $t_\text{n}$ represents the time when the clock signal arrives at memory cell n. The total clock skew of a circular shift register (dotted) is equal to zero. (b) Diagram of two types of comb filters: feedback [left] and feedforward [right], both with a delay and scaling factor stage. (c) Block diagram of a DMTG, using a N-bit CSR and a single stage comb filter step. Full symmetric clock distribution is used with N/2 bits having concurrent flow and the remaining N/2 having counter flow of data and clock. A non destructive readout cell (NDRO) is used to ensure a way of writing data in the CSR and forming the loop to generate the pulse train, as well as used in the comb filter to allow a zero delay output (Set and Reset). D-JTL (delay JTL) is used as a tuneable way of changing the fluxon propagation speed in the JTL.}
\label{fig:fig4}
\end{figure}

CSRs are an extension of linear shift registers, where the output of the last memory cell is connected to the input of the first, creating a circular flow of data (Fig. \hyperref[fig:fig4]{4(a)}). An N-bit CSR consists of N D Flip-Flops (DFFs) connected in series using Josephson Transmission Lines (JTLs), to control the delay between cells. To clock the CSR, 2-way splitters are usually used, however, in this device we use 3-way splitters to meet the requirements of our clocking network, as will be explained below.

Shift registers are synchronous circuits, meaning they require a clock to advance to the next internal state, independent of the data inputs. The type and architecture of the clock distribution network is therefore crucial for the correct operation of this device, considering all internal and external delays on the data and clock paths. Particularly important is the clock skew, $t_{\text{cs}}$, of the circuit, which for a CSR must equal to 0 \cite{RN13,RN19}. This quantity is defined as the time between a clock pulse arriving at the $i$th memory cell and the same clock pulse reaching the $(i+1)$th cell (see Fig. \hyperref[fig:fig4]{4(a)}).  For a CSR:

\begin{align}
    t_{\text{cs}}^{\text{Total}} = \sum_{i=1}^{\text{N}} t_{\text{cs}}^i = (t_1 - t_2) + (t_2-t_3) + \ldots + (t_N - t_1) = 0
\end{align}
where $t_{\text{cs}}^i = (t_i - t_{i+1})$, and $t_i$ is the time when clock pulse arrives at the $i$th cell. To avoid any racing conditions, the clock skew for each cell plus the data delay between cells must be greater than the hold time of the cell. This must still be met, even though the total clock skew of the CSR is zero. 

To satisfy this condition, a type of clocking network was designed which is based on a type of symmetrical-mixed clock design as seen in Refs. \cite{AUTO1,RN13,RN25}. Here, N/2 bits of the shift register are clocked in a concurrent way and the remaining N/2 bits are clocked in a counter flow manner, but only one input and output of data are used, as shown in Fig. \hyperref[fig:fig4]{4(c)}. This ensures that the two blocks have negative and positive clock skews, respectively, but when combined in a loop the total equals to zero. Another alternative is to use a standard binary tree clock distribution, which has a higher overhead of splitters. For binary-tree clocking, the clock skew between adjacent memory cells is zero, so the sum will always also be zero.
The total amount of splitters necessary for this type of clock network is $\text{N}-1$, whereas for the symmetrical type it requires $\text{N}/2$.


\begin{figure*}[ht]
    \centering
    \resizebox{\textwidth}{!}{%
        \begin{tikzpicture}
            \node[anchor=center] (sub1) at (0,0) {\includegraphics[]{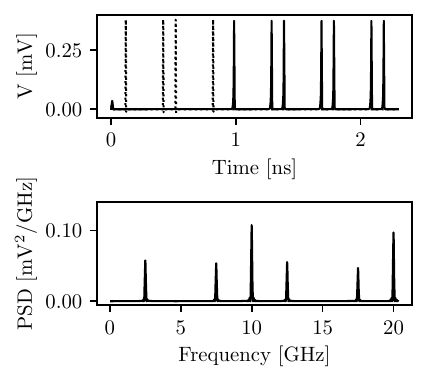}};
            \node[anchor=center] (sub2) at (7.2,0) {\includegraphics[]{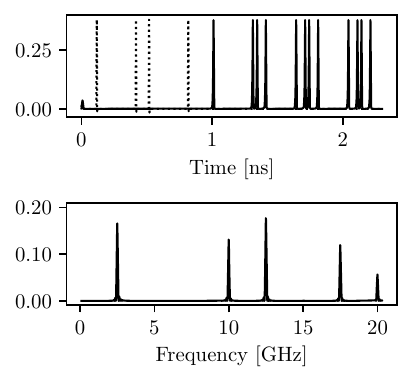}};  
            \node[anchor=center] (sub3) at (12.5,0.28) {\includegraphics[]{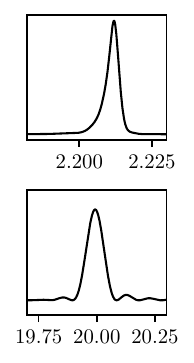}};
            \node[text width=4cm,align=center] at ($(sub1.center)+(0.6,3.5)$) {Pattern `1001 1001'};
            \node[text width=4cm,align=center] at ($(sub2.center)+(0.6,3.55)$) {Pattern `1001 1001'\\\& {Comb Filter}};
            \draw[thick,dashed,color=denim,fill=denim,fill opacity=0.1] ($(sub2.south east)+(-0.9,1.5)$) rectangle ++(0.4,0.8) ;
            \draw[thick,dashed,color=denim,fill=denim, fill opacity=0.1] ($(sub2.north east)+(-0.9,-2)$) rectangle ++(0.25,1.6) ;
            \draw[thick,-{Triangle[length=5pt]},dashed] ($(sub2.south east)+(-0.65,5.5)$) -- ++(32pt,0);
            \draw[thick,-{Triangle[length=5pt]},dashed] ($(sub2.south east)+(-0.50,1.9)$) -- ++(28pt,0);
            \node[align=center] at ($(sub1.north west)+(0.3,0.3)$) {\textbf{(a)}};
            \node[align=center] at ($(sub2.north west)+(0.3,0.3)$) {\textbf{(b)}};
        \end{tikzpicture}
    }
\caption{Frequency spectra obtained from PSCAN2 time simulations using the Lomb-Scargle periodogram algorithm for a 5 ns pulse train. The same pattern used in Fig. \hyperref[fig:fig1]{1(c)} is used for comparison, without (a) and with (b) the comb filtering stage set very close to the 0.333 ns delay to obtain the best separation between the two tones, 2.5 and 7.5 GHz (as shown in Fig. \ref{fig:fig3}). Dashed lines on time domain plots show the input data (pre-loading) sent to the CSR. Insets highlight one SFQ pulse (top), as well as the width of one of the frequency tones (bottom).}
\label{fig:fig5}
\end{figure*}
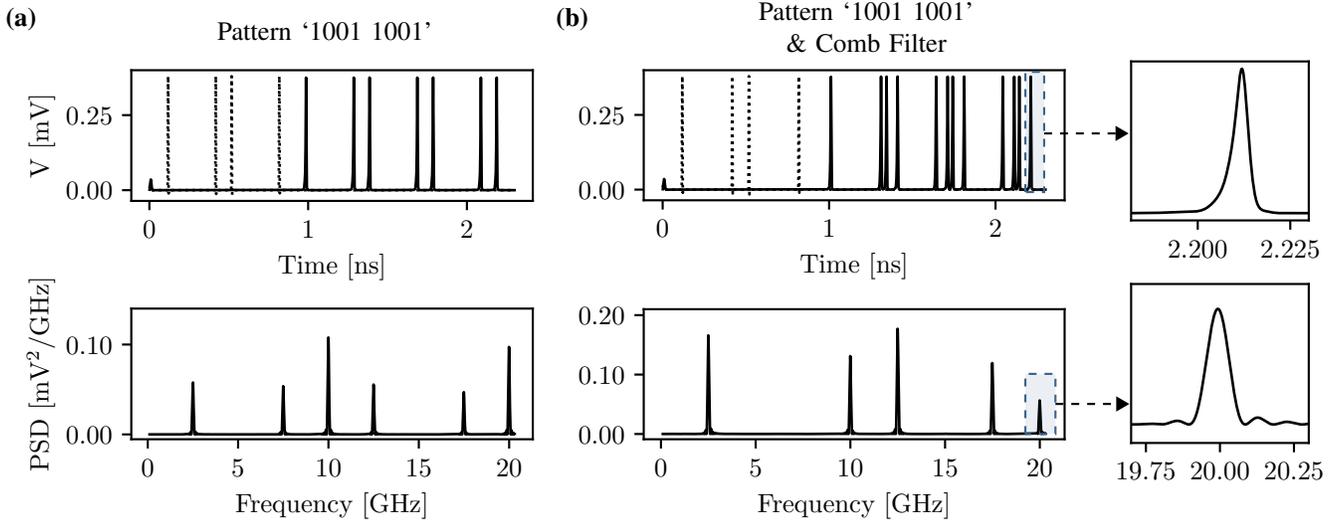
\subsection{Comb Filtering}
Comb filtering in RSFQ circuits must be modified slightly compared to the standard digital signal processing filter, since SFQ pulses cannot be summed in amplitude like normal signals can. This means that the two signals generated must be merged, instead of summed. Also, the scaling factor seen in Eq. \eqref{eq:8} is harder to achieve, specifically when $\alpha < 1$, since the amplitudes of SFQ pulses are set during the design phase by the critical current density $j_c$ and I$_\text{c}$R$_\text{N}$ product of the junctions. It is also possible to use an amplifying JTL to increase the voltage of the SFQ pulse and obtain $\alpha > 1$, which can be useful for increasing the power of the output tones. 

The implementation of comb filtering in this work is based on the feedforward architecture with a unity scaling factor. The circuit has three basic RSFQ cells: one splitter to generate the two data paths, one merger to combine the two pulses and one long JTL, which we call a delay JTL, which delays the propagation of SFQ pulses in one branch of the circuit. This delay ($\tau$) is tuneable by changing the bias current of the entire delay JTL, based on fluxon propagation and interaction in JTLs \cite{RN43}. This implementation is shown in Fig. \hyperref[fig:fig4]{4(c)}, alongside a full block diagram of the multi-tone generator device, comprised of a N-bit CSR and a single stage feedforward comb filtering stage, with one data input and one data output. A NDRO cell is used to break the loop of the CSR and choose whether the CSR is in write (SET=0) or read mode (SET=1), which opens and closes the loop,  respectively. Another NDRO switch is added to the comb filter to allow for an output with no delayed pulse. Alternatively, an RSFQ DC switch could be used as a replacement for NDRO cells, to reduce the complexity and overhead of the design.


\section{Simulations} \label{ref:sec4}

To test the operation of the device, a simulation model was implemented using PSCAN2 superconductor circuit simulator \cite{RN11}. A netlist of the device was designed using SeeQC's high density fabrication process parameters, specifically devised for digital circuits and quantum applications \cite{FAB1}. A modular approach is used to optimise the design, starting with the memory cells and the shift register and further adding the comb filter stage. The optimisation process was done at 10GHz (a conventional operation frequency for superconducting quantum circuits), therefore for higher frequencies the margins are expected to be narrower. The global margins XI, XJ, and XL, represent the total deviations of the bias currents, junction critical currents, and inductances, respectively, in the device, such that the correct operation is not compromised. Maximum margins are capped at 40\%. These are obtained from the full circuits and tested at different clock frequencies (namely, 10, 20 and 50 GHz) and with different patterns on the CSR. For operation around 50 GHz and above, slight modifications of the CSR were made, increasing the I$_\text{c}$R$_\text{N}$ product for faster practical circuit operation and removing some JTLs between data cells. A maximum frequency of 86 GHz was obtained with correct operation of the CSR without the comb filter, as a demonstration of the robustness of the design. A summary of these tests is shown in Table \ref{tab:tab1}. 

\renewcommand{\arraystretch}{1.2}
\begin{table}[]
    \centering
    \begin{tabular}{c|ccc|ccc}
    & \multicolumn{3}{c|}{Pattern `1000 1000'} & \multicolumn{3}{c}{Pattern `1001 1001'} \\ \cline{1-7} \rule{0pt}{2ex} 
    & 10 GHz & 20 GHz & 50 GHz$^*$ & 10 GHz & 20 GHz & 50 GHz$^*$  \\ \cline{1-7} \rule{0pt}{3ex} 
        XI & [-28,25] & [-21,25] & [-21,18] & [-28,25] & [-22,23] & [-21,18] \\ \rule{0pt}{2ex}
        XJ & [-22,25] & [-22,18] &  [-16,23] & [-20,25] & [-19,18] & [-16,18] \\ \rule{0pt}{2ex}
        XL & [-40,40] & [-40,40] &  [-40,40] & [-40,40] & [-40,40]  & [-35,34]
    \end{tabular}
    \caption{Simulated Global Margins for an 8-bit device (in percentages).}
    \label{tab:tab1}
\end{table}

The output frequency spectrum was obtained from the simulations by using the Lomb-Scargle periodogram \cite{RN44}. This algorithm provides an estimation of the discrete Fourier Transform for samples with unevenly spaced time data, which is crucial for the analysis of PSCAN2 simulations due to its dynamic time step.

\begin{figure}[!ht]
\begin{subfigure}{\columnwidth}
    \resizebox{\columnwidth}{!}{%
        \begin{tikzpicture}
            \node (sub1) at (0,0) {\includegraphics[]{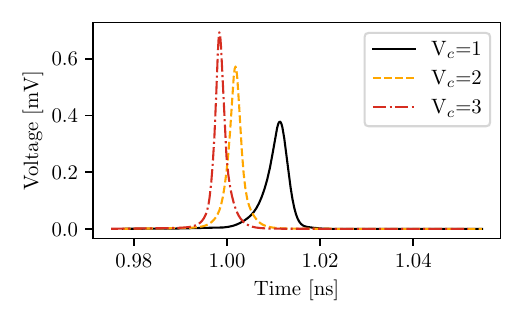}};
            \node[align=center] at (sub1.north west) {\textbf{(a)}};
        \end{tikzpicture}
    }
\end{subfigure}
\par
\begin{subfigure}{\columnwidth}
    \resizebox{\columnwidth}{!}{%
        \begin{tikzpicture}
            \node (sub1) at (0,0) {\includegraphics[]{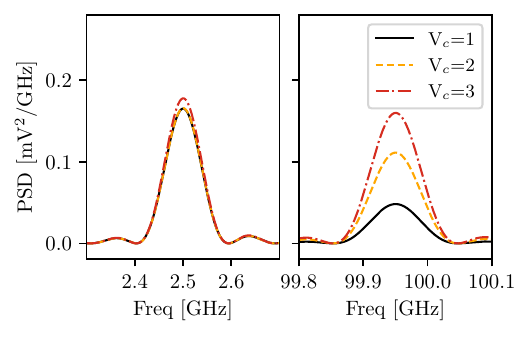}};
            \node[align=center] at (sub1.north west) {\textbf{(b)}};
        \end{tikzpicture}
    }
\end{subfigure}
\par
\begin{subfigure}{\columnwidth}
    \resizebox{\columnwidth}{!}{%
        \begin{tikzpicture}
            \node (sub1) at (0,0) {\includegraphics[]{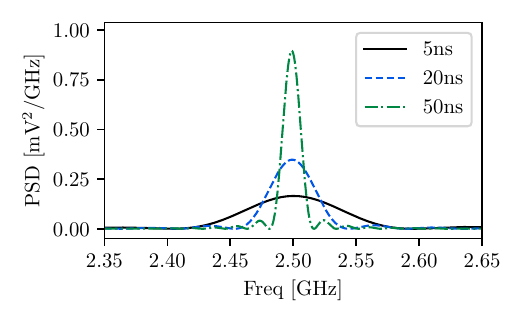}};
            \node[align=center] at (sub1.north west) {\textbf{(c)}};
        \end{tikzpicture}
    }
\end{subfigure}
\caption{Influence of pulse train characteristics on frequency spectrum for the same pattern applied to an 8-bit CSR, as seen in Fig. \ref{fig:fig5}, focusing on the 2.5 GHz tone. (a) An increasing characteristic voltage, V$_c$=I$_\text{c}$R$_\text{N}$, of the output junctions in the circuit, results in a narrower pulse with an increased amplitude. Difference in time shown corresponds to the small jitter in PSCAN2 simulations. (b) Increasing V$_c$, the output power of lower frequency tones remains approximately constant, while higher frequencies are amplified. (c) To decrease the width of each tone, a longer pulse train should be considered. For a 50 ns pulse train, a bandwidth of 15 MHz is obtained, compared to 84 MHz for a 5 ns pulse train. V$_c$ is in dimensionless PSCAN2 units, where unity equals 0.287 mV.}
\label{fig:fig6}
\end{figure}

The dependence of the frequency spectrum on the characteristics of the SFQ pulses comprising the pulse train was also investigated and is shown in Fig. \ref{fig:fig6}. Firstly, by changing the characteristic voltage V$_c$ of the output junctions, we observe a change in the relative powers between the tones of the signal, as shown in Fig. \hyperref[fig:fig6]{6(b)}. This is in agreement with the result obtained in Ref.\cite{RN31}, where for narrow pulses, the higher frequency components have increased power. This can become a problem when working with superconducting circuits since at some critical power these high frequency tones may break Cooper pairs and poison the system with quasiparticles \cite{RE1}. In practice, the advantage of having narrower pulses is to make the pulse trains and subsequently the control pulses, shorter in time. Since the area of the pulse, $\Phi_0$, remains the same, the average power output also does not change. 

Increasing the pulse train length improves the bandwidth of the output tones, as seen in Fig. \hyperref[fig:fig6]{6(c)}. The longer the pulse train, the narrower the tones, which in practice means our control pulse trains cannot be too short in time. The average power output is the same in the time domain (since the peak power and the duty cycle remain constant), although there is more energy in the system.

Clock jitter is another metric which exists in real experiments and can affect the frequency spectrum of the output signal. While the base simulations shown do not include a known and controllable source of jitter, there is a small variation of the clock period due to the way the simulation, particularly the dynamic time step calculation, is performed, but accounts for less than few picoseconds. Adding more jitter, the tones tend to shift from their expected value by a few MHz (as seen on the right side of Fig. \hyperref[fig:fig6]{6(b)}, where the tone at 100 GHz slightly shifts) and noise at higher frequencies increases.

When discussing powers with pulse trains and frequency spectra, one must distinguish between a couple of variables. The average power in the time domain of a pulse train will be given by:
\begin{align}
    \text{P}_\text{avg} = \text{P}_\text{peak} \cdot \text{Duty Cycle} = \text{P}_\text{peak} \cdot \frac{\sigma_{\text{pulse}} }{\text{T}_{\text{clk}}}\cdot \frac{\text{n}}{\text{N} }
\end{align}
where T$_\text{clk} =  1 / f_{\text{clk}}$, $\sigma_{\text{pulse}}$ is the width of each SFQ pulse and n is the number of set bits in an N-bit pattern. Increasing the pulse length, as done in this work, increases the total energy in the signal, but the average power in time domain remains. For the Lomb Scargle periodogram algorithm, the result is a metric of the squared amplitudes of each Fourier component at each frequency, with units of mV$^2$ in this work. This is known as the Power Spectrum (PS). Since we are dealing with discrete signals, that is, sampling rate and number of points are finite, one drawback is that the DFT and all estimations scale with the number of points N of our original signal, making the amplitude of the calculation arbitrary and with no real meaning. For the transform to be useful, we calculate the Power Spectral Density (PSD) which is obtained by dividing the power spectrum by the frequency resolution of the spectrum:
\begin{align}
    \text{PSD} \equiv \frac{\text{PS}}{f_s \text{N}}
\end{align}
where $f_s$ is the frequency resolution and N is the number of sample points. Using the PSD, the area under each tone represents the fraction of the signal power that is concentrated around that tone. Increasing the total length of the pulse train therefore decreases the tone width, but amplitude increases so that the total area remains constant, since the input power remains the same (higher length means more energy but averaged over longer period).

\section{Outlook} \label{ref:sec5}
The DMTG device presented and discussed herein has applications in both classical RSFQ circuitry, as well as a control circuit for quantum systems that also require a cryogenic environment and the higher frequencies achieved with RSFQ and ERSFQ technologies. Two examples which motivated this work are qubit and SNSPD control and readout systems. 

For standard qubit control, a microwave pulse is carefully crafted to excite single qubit gates and drive qubit state rotation on the XX and YY axes. For this method, our device could be used as a way to multiplex the control and use a single line for many qubits, after it is filtered and impedance matched on a RF module, as shown in Ref. \cite{RN29}. 

Currently, SFQ-based digital control is achieved using a simple DC-SFQ converter (or a combination of such) and clocked at a subharmonic of the qubit transition frequency \cite{RE1}. The idea is that each pulse rotates the qubit state slightly around the XX or YY axis of the Bloch Sphere, and by controlling the number of pulses the desired final state can be obtained. The subhamornic is used to avoid driving the qubit directly, since the SFQ system is capacitively coupled to it.  Our device could be used like this in a fully digital way or can be used to implement more complex pulse sequences as shown in Ref. \cite{RN99}. These sequences can optimise SFQ-based qubit control by reducing leakage outside the computational space and increasing the fidelity of single qubit gates.

\begin{figure}[!ht]
    \begin{tikzpicture}
        \input{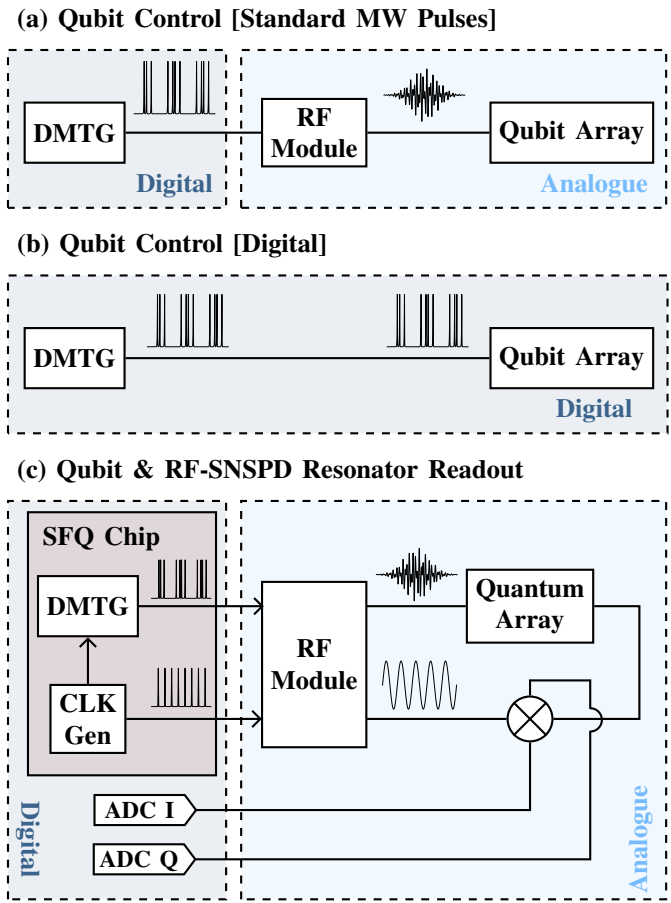}
    \end{tikzpicture}
    \caption{Proposed SFQ-Based Architectures for Qubit and SNSPDs applications utilising the Digital Multi-Tone Generator (DMTG) device presented in this work. (a) Qubit control using a RF module (see Ref. \cite{RN29}) to filter the complex pulse train into an analogue signal to drive XX and YY qubit state rotations. (b) Directly applying the SFQ pulses to the qubit yields a full digital control method, also working as XX and YY qubit gates (see Ref. \cite{RE1,RN99}). (c) Resonator readout is also possible using a heterodyne scheme where both the multi tone signal and a local oscillator are generated on-chip. Down-conversion can be possibly substituted by a RSFQ based ADC to directly digitise the signal from the quantum system, creating a full RSFQ based control system.}
    \label{fig:fig7}
\end{figure}

Superconducting Nanowire Single Photon Detectors (SNSPD) are another area where superconductor electronics are foreseen to improve scalability. In particular, large arrays of RF-SNSPDs (SNSPDs integrated in lumped element resonators) can be used in conjunction with RSFQ electronics to create a control and readout system that can scale more efficiently than DC-SNSPDs \cite{RN38}. The output signal with multiple tones would be sent to an array of resonators that are coupled to the SNSPDs, and, using frequency multiplexing, all pixels could be probed simultaneously with only one input. The readout architecture would closely follow the same one used for microwave electronics, where a multi-tone input signal is sent to the signal, filtered, and down-converted using IQ mixers, to then be detected by an Analogue-to-Digital Converter, which could also be implemented with RSFQ circuitry \cite{RN46,RN47}. In this case, no down-conversion would be necessary. Additionally, time domain multiplexing could be used to sweep the local oscillator signal (LO) such that the down-converted signal frequency could always fit within the bandwidth of the ADC, reducing the sample rate, but increasing the total bandwidth available for the resonators.
The same architecture could be implemented to probe an array of readout resonators coupled to qubits in the standard dispersive regime \cite{DR1}.
A summary of these architectures is shown in Fig. \ref{fig:fig7}.

\section{Conclusion} \label{ref:sec6}
In conclusion, we have presented a novel RSFQ device that uses the properties of pulse trains and their Fourier Transform, to create multi-tone signals that depend on the data (pattern) loaded in the memory cells of an N-bit CSR. We have shown the existence of and the means to obtain all unique patterns for an N-bit CSR, although the search for optimal and unique patterns is computationally complex and requires exponential time. Together with a feedforward comb filter stage, the tones generated are further concentrated around certain frequencies, which increases the tuneability of the output, as well as increasing the amplitude of the desired tones. Additionally, stacking comb filters can further improve the amplitude and spectrum.

In terms of components, the device is relatively simple to fabricate, uses well developed and studied RSFQ components, and the number of junctions in the circuit grows reasonably well with increasing N. The simulated circuits show a great match with the predicted frequency spectra for every pattern and it is also shown that changing the SFQ pulse width, pulse train length, and clock periodicity has an influence on the result. This is all consistent with the results predicted theoretically. The average margins are shown to be quite good for different patterns at around 20\%, which is similar to margins obtained experimentally for RSFQ devices with complexity involving few hundred junctions. It is also shown that the maximum operating frequency is dependent on the characteristic voltage of the junctions, but with this fabrication parameters, operation close to 100 GHz can be achieved.

In the future, a way of improving the algorithm to search for unique patterns with an improved time complexity is required, as well as a method to estimate a set of patterns which can be used, given the tones desired by the user.

\section*{Acknowledgement}
This work was funded and supported by the National Manufacturing Institute Scotland and the Scottish Research Partnership in Engineering (SRPe) through the Industrial Doctorate Programme NMIS-IDP/025, SeeQC UK Limited and the University of Glasgow through EPSRC grant no. EP/T025743/1.
\bibliographystyle{IEEEtran}
\bibliography{IEEEabrv,refies}
\end{document}